\documentstyle[prl,twocolumn,aps,graphicx]{revtex}

\begin{document}

\twocolumn[\hsize\textwidth\columnwidth\hsize\csname    
@twocolumnfalse\endcsname                               

\begin{title} {\Large \bf
Spin Ice State in Frustrated Magnetic Pyrochlore Materials}
\end{title}

\author{Steven T. Bramwell$^{1}$ and Michel J.P. Gingras$^{2,3}$}
 
\address{$^{1}$Department of Chemistry, University College London, 20 Gordon
Street, London, WC1H OAJ, United Kingdom}
\address{$^{2}$Department of Physics, University of Waterloo, Ontario, Canada
N2L 3G1}
\address{$^{3}$Canadian Institute for Advanced Research, 180 Dundas Street W.,
Toronto, Ontario, Canada M5G 1Z8}
\date{\today}
\maketitle
\begin{abstract}
A frustrated system is one whose symmetry precludes the
possibility that every pairwise interaction (``bond'')
in the system can be satisfied at the same time. Such systems are
common in all areas of physical and biological science. In the
most extreme cases they can have
a disordered ground state with ``macroscopic'' degeneracy, that is, one
that comprises a huge number of equivalent states of the same energy.
Pauling's description of the low temperature
proton disorder in water ice was perhaps the first recognition of this
phenomenon, and remains the paradigm. In recent years a new class of
magnetic
substance has been characterised, in which the disorder of the
magnetic moments at low temperatures is precisely
analogous to the proton disorder in water ice. These substances, known
as spin ice materials, are perhaps the ``cleanest'' examples of such highly
frustrated systems yet discovered. They offer an unparalleled opportunity
for the study of frustration in magnetic systems
at both an experimental and a theoretical level.
This article describes the essential physics of spin ice,
as it is currently understood,
and identifies new avenues for future research on
related materials and models.

\vspace{-5mm}
\end{abstract}

\vskip2pc]                                              


\narrowtext

Competing or frustrated interactions are a common feature of condensed
matter systems.
Broadly speaking, frustration arises when a system cannot, due to local
geometric constraints, minimize all
the pairwise interactions simultaneously (\cite{Toulouse}).
In some cases, the frustration can be so intense that
it induces novel and complex phenomena. Frustration is at the origin of
the intricate structure of molecular crystals, various phase
transitions in liquid crystals and the magnetic
domain structures in ferromagnetic films.
It has also been argued to
be involved in the formation of the stripe-like structures observed in cuprate
high-temperature superconductors. The concept of frustration is
a broad one that extends beyond the field of condensed matter physics.
For example,
the ability of naturally occurring
systems to ``resolve''
frustrated interactions has been argued to have
bearings on life itself, exemplified by the folding of a
protein to form in a single and well prescribed structure with
biological functionality.

Historically, the first frustrated system identified was
crystalline ice, which has residual frozen-in disorder remaining down
to extremely low temperature, a property known as
residual, or zero point
entropy.  In 1933, Giauque and
co-workers accurately measured this entropy~(\cite{Giauque,Giauque2}),
enabling Pauling to offer his now famous
explanation in terms of the mismatch between the crystal symmetry and
the local bonding requirements of the water molecule~(\cite{Pauling}).
He predicted a special type of proton disorder that obeys the so-called ``ice
rules''.
These rules, previously proposed by Bernal and Fowler~(\cite{Bernal}),
require that two protons are near to
and two are further away from each oxide ion, such that the crystal structure
consists of hydrogen bonded
water molecules, H$_2$O(see Fig.1).
Pauling showed that the ice rules do not lead to order in the proton
arrangement but rather, the ice ground state is ``macroscopically
degenerate''. That is to say, the number of degenerate, or energetically
equivalent proton arrangements diverges exponentially with the number $N$
of water molecules. Pauling estimated the degeneracy to be $\sim
(3/2)^{N/2}$ where $N$ is the number of water molecules, typically $\sim
10^{24}$ in a macroscopic sample. This leads to a disordered ground state
with a measurable zero point entropy $S_0$ related to the degeneracy:
$S_0 \sim$ $(R/2)\ln(3/2)$, where $R$ is the molar gas constant.  Pauling's
estimate of $S_0$ is very close to the
most accurate modern estimate~(\cite{Nagle}) and consistent with
experiment~(\cite{Giauque}).
The disordered ice-rules proton arrangement in water ice
was eventually confirmed by neutron diffraction
experiments~(\cite{Wollan,Li}).

Magnetic systems offer themselves as the ideal benchmark for generic
concepts pertaining to collective phenomena in nature. This is due
in part to
the availability of a large variety of diverse magnetic materials that
can be chosen to approximate simple theoretical ``toy models'' of
collective behaviour, and in part
to their ease of study by a battery of
experimental techniques. Over the last fifty years, experimentalists
have characterised new classes of frustrated magnetic behaviour and
theoreticians have been motivated by the broad conceptual applicability
of magnetic models to investigate simple frustrated spin
systems (\cite{frus1,frus2,frus3}). These
include ``energetic'' generalisations of the ice model that display a
wealth of interesting thermodynamic phenomena in close resemblance with
those observed in real ice~(\cite{Baxter,Lieb}).
However, while theoretical studies of
ice-like phenomena in frustrated ice models have long flourished, very
few, if any real magnets could be found to display a close
thermodynamic resemblance to common ice.
This remained for sometime a disappointing
situation where close contact between theoretical studies on
magnetic ice models and real systems was lacking, a somewhat
untenable predicament in science where one is generally aiming at
testing theoretical concepts against experiments, and vice versa.

\vspace{-4cm}
\begin{figure}
\begin{center}
\rotatebox{0}{\includegraphics[width=9cm]{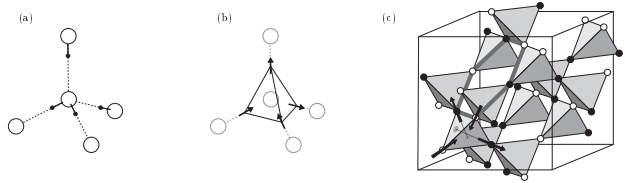}}
\vspace{-5cm}
\caption{
 (A) Local proton arrangement in water ice, showing oxide ions (large white
circles) and protons (hydrogen ions, small black circles). Each oxide is
tetrahedrally coordinated with four other oxides, with two near
covalently bonded protons, and two are further
hydrogen bonded protons.  The low energy configurations obey the
so-called ``ice rules'' 
([5])
 where each oxide has two
``near'' and two ``far'' protons. (B) Same as in (A), but where now the
position of the protons are represented by displacement vectors
(arrows) located at the mid-points
of the oxide-oxide lines of contact.  The
ice rules in (A) translates into a ``two in $-$ two out'' configuration
of the displacement vectors.  (C) Pyrochlore lattice of corner-sharing
tetrahedra, as occupied by the magnetic rare earth ions
in the spin ice materials Ho$_2$Ti$_2$O$_7$ and Dy$_2$Ti$_2$O$_7$.
The magnetic Ising moments
occupy the corners of the
tetrahedra, as shown on the lower left ``downward'' tetrahedron of the
lattice (arrows). The spins here are the equivalents of the
proton displacement vectors in (B).  Each spin axis is along the
local $\langle  111 \rangle $ quantization axis, which goes from one
site to the middle of the opposing triangular face (as shown by the
disks) and meets with the three other $\langle 111 \rangle $  axes in
the middle of the tetrahedron. In the spin ice
materials the ``two in $-$ two-out''
condition arises from the combined effect of magnetic coupling and
anisotropy. For clarity, other spins on the lattice are denoted by
black and white circles, where white represents a spin pointing into a downward
tetrahedron while black is the opposite. The entire lattice is shown
in an ice-rules state (two black and two white sites for every
tetrahedron).
The hexagon (thick gray line) relates to the discussion in section III.
It shows the smallest
possible loop move involving multiple spins, and corresponds to
reversing all colors (spins) on the loop to produce a new ice-rules
state.  These extended type of excitations or processes
are the ones that allow the system to explore the quasi-degenerate ice rule
manifold at low temperature.
Common water ice at atmospheric pressure, ice I$_h$, possess a hexagonal
structure while here, the magnetic lattice has cubic symmetry.
Strictly speaking, the Ising pyrochlore problem is equivalent to
cubic ice, and not the hexagonal phase. Yet, this
does not modify the ``ice-rule'' analogy (or mapping) or the connection between
the statistical mechanics of the local proton coordination in water ice
and the low temperature spin structure of the spin ice materials.
}
\end{center}
 \end{figure}
\vspace{1mm}

Anderson had noticed in 1956 the formal analogy that
exists between the statistical mechanics of cation
ordering on the cubic B-site lattice in ``inverse'' spinel
materials and the statistical mechanics
of antiferromagnetically coupled
two-state Ising magnetic moments on the same lattice (referred
to here as the pyrochlore lattice, Fig.1C)~(\cite{Anderson}).
Both systems were shown to
map exactly onto Pauling's model of proton disorder in ice.
The realization of Anderson's model in an antiferromagnetic
material would require spins to
point along or antiparallel to a
global $z$ axis direction. However,
there is no reason to prefer the $z$
over the $x$ or the $y$ direction in a lattice
with global cubic symmetry, and this renders
the global antiferromagnetic Ising model
unrealistic with no direct relationship to any real magnetic material.
The experimental situation changed in 1997 when it was noticed by
Harris {\it et al.}~(\cite{PRL1}) that
a model of ferromagnetism on the pyrochlore lattice
would exactly map onto the ice model so long
as each Ising-like magnetic moment was constrained
to point along the axis joining the
centers of the two tetrahedra to which it belongs (Fig. 1C).
This was
a surprising observation, because naively one would not
expect frustration in a ferromagnet.
However, the ferromagnetic model is
compatible with cubic symmetry and was observed to
be approximated by the apparently ferromagnetic pyrochlore material
Ho$_2$Ti$_2$O$_7$~(\cite{PRL1}).
This constituted the first simple physical
realization of a real three dimensional magnetic analogue of
common ice, and the name ``spin ice'' was coined to
emphasize this analogy.

Experiments on spin ice have mirrored, using modern sophistication,
those originally conducted on water ice. However, the spin ice
materials lend themselves more readily to experiment than does water
ice and more closely approximate tractable theoretical models.
This has led to much recent interest devoted to the problem of zero point
entropy and to the study of the
broad consequences of geometric frustration.
We review the recent experimental and theoretical developments in the study of
spin ice materials, and discuss what are possible new and exciting
avenues of research in this problem.

\section{Discovery of Spin Ice}

In a flux grown crystal of
Ho$_2$Ti$_2$O$_7$~(\cite{Godfrey}) (Fig. 2) the
octahedral habit reflects the cubic symmetry of the pyrochlore
structure while the amber colour and strong reflectivity are indicative
of a band gap near the visible/ultraviolet boundary (3.2 eV).
In Ho$_2$Ti$_2$O$_7$, the Ho$^{3+}$ ions occupy a pyrochlore lattice
of corner-linked tetrahedra (illustrated in Fig. 1C). Magnetism arises from
the Ho$^{3+}$ ions, Ti$^{4+}$ being
non-magnetic.  Ho$^{3+}$ has a particularly large magnetic moment of
approximately
10$\mu_{\rm B}$ that persists to the lowest temperatures and makes the
crystals sufficiently paramagnetic to stick to a permanent magnet even
at room temperature (Fig. 2).  The large, temperature independent,
moment is ensured by the local crystallographic environment of Ho$^{3+}$ in
the pyrochlore stucture~(\cite{Blote,Mamsurova,Bramwell,Jana,Rosenkranz}).
Each tetrahedron of Ho$^{3+}$ ions has an oxide ion at its centre, so two
of these oxide ions lie close to each Ho$^{3+}$ along the $\langle
111\rangle$ crystallographic axis that connects the centre of the
tetrahedron to its vertex. The anisotropic crystallographic environment
changes the quantum ground state of Ho$^{3+}$ such that its magnetic moment
vector has its maximum possible magnitude and lies parallel to the local
$\langle 111\rangle$ axis. In the language of quantum mechanics the $^5I_8$
free ion state is split by the local trigonal crystal field such that the
ground state is
an almost pure $|J, M_J \rangle =|8,\pm 8 \rangle$ doublet with
$\langle 111\rangle$ quantization axis. The first excited state lies
several hundreds of Kelvin above the ground state as revealed by inelastic
neutron scattering measurements~(\cite{Rosenkranz}). At temperatures of
the order of ten Kelvin or below, the excited states are not accessed
thermally.  The
Ho$^{3+}$ moments therefore behave as almost
pure two-state spins that approximate classical
Ising spins pointing ``in'' or ``out'' of the elementary tetrahedra
(Fig. 1). Direct evidence of the single ion anisotropy comes from bulk
magnetization (\cite{Mamsurova,Bramwell,Cornelius}), where saturation is
observed at roughly half the expected value, owing to the fact that
applied fields of several Tesla are too weak to turn the Ho$^{3+}$
significantly away from their local quantisation axes.

\begin{figure}
\begin{center}
\rotatebox{0}{\includegraphics[width=6cm]{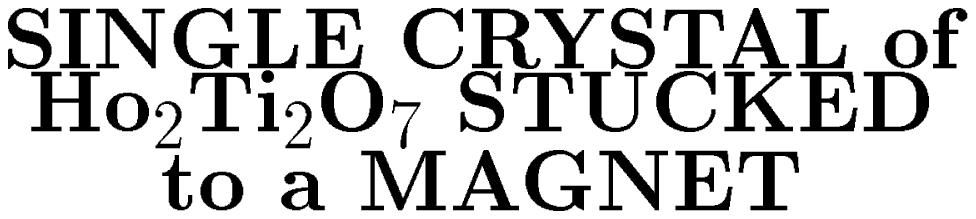}}
\vspace{-5cm}
\caption{Flux-grown octahedral crystal of Ho$_2$Ti$_2$O$_7$ stuck to
a NdFeB permanent magnet at room temperature. The strong paramagnetism
reflects the large magnetic moment of Ho$^{3+}$.}
 \label{phase}
\end{center}
 \end{figure}
Experimental investigation of the spin correlations in
Ho$_2$Ti$_2$O$_7$ began in 1996 (\cite{PRL1}).  Prior to this,
susceptibility
measurements (\cite{Cashion}) had
revealed a peak in the susceptibility
of Ho$_2$Ti$_2$O$_7$ at $\sim 1$ K, suggestive of antiferromagnetic
interactions.
The first muon spin relaxation ($\mu$SR) and neutron scattering
experiments seemed to confirm this frustrated antiferromagnetic scenario
with no evidence of long
range order down to $\sim 50$ mK~(\cite{PRL1,HarrisMMM}).
However, a pyrochlore antiferromagnet,
with essentially infinite local $\langle 111\rangle$ Ising anisotropy,
should develop a
long range ordered state at
a critical temperature of
order of the Curie-Weiss temperature, $\sim 1$ K,
(\cite{JPCM,infinite-ising}).
Consequently,
the failure of Ho$_2$Ti$_2$O$_7$ to display a transition down to
$\sim 50$ mK was found to be rather paradoxical.
However, new susceptibility studies
soon suggested a rather different picture.
The large moment of Ho$^{3+}$ was found
to produce a strong demagnetizing field that caused the experimental
Curie-Weiss temperature (measured by the intercept of the inverse
susceptibility versus temperature curve) to be either ferromagnetic or
antiferromagnetic depending on crystal shape.  Careful correction for
this shape-dependent effect indicated that $\theta_{\rm CW} = 1.9 \pm 0.1$
K, an intrinsically ferromagnetic value. It therefore seemed that
Ho$_2$Ti$_2$O$_7$ should be described, at least to a first
approximation, as a $\langle 111\rangle$ Ising ferromagnet. But this
description at first seemed contrary to the observed absence of long
range order $-$ it was ``obvious'' that a ferromagnet should order at low
temperature!

As often in science, the paradox was resolved as soon as the
``obvious'' was abandoned in the face of experimental evidence.
Calculation showed that the ground state of a tetrahedron
of ferromagnetically coupled $\langle 111\rangle$ Ising spins is the
``two in $-$ two out'' state illustrated in Fig 1. It was then recalled
that Anderson had shown the pyrochlore lattice to be the medial lattice
(lattice formed by the mid points of the bonds) of the diamond-like
oxide lattice of cubic ice (\cite{Anderson}). Hence, the ``two in $-$ two
out'' condition is analogous to the ice rules and the ground state of
the nearest neighbour ferromagnetic model is, like that of ice,
macroscopically degenerate~(\cite{PRL1,JPCM}).
The absence of long range order in
Ho$_2$Ti$_2$O$_7$ could then be explained at a qualitative level and
the ``spin ice'' model, the $\langle 111\rangle$ Ising ferromagnet,
was christened.
This simple model was found to be consistent with the field-induced
ordering patterns observed by neutron scattering (\cite{PRL1}). On the
basis of similar susceptibility properties, Dy$_2$Ti$_2$O$_7$ and
Yb$_2$Ti$_2$O$_7$ were also suggested to be spin ice materials
(\cite{PRL2}). So far, only Ho$_2$Ti$_2$O$_7$~(\cite{PRL1,Bramwell-Sq}),
Dy$_2$Ti$_2$O$_7$~(\cite{ramirez-nature,Fennell}),
and more recently Ho$_2$Sn$_2$O$_7$~(\cite{Kadowaki}),
have been positively confirmed.

The magnetization and elastic neutron scattering
measurements described above provided the
initial compelling arguments
for the spin ice phenomenology associated
with Ho$_2$Ti$_2$O$_7$~(\cite{PRL1}). However, specific heat
measurements by Ramirez {\it et al.} (\cite{ramirez-nature,HarrisNature}) on
Dy$_2$Ti$_2$O$_7$ have given a
more direct experimental evidence of the
macroscopic degeneracy associated with the spin ice rules.
The top panel of Fig. 3 shows the temperature dependence of
the magnetic specific heat, $C(T)$, for a powdered sample of
Dy$_2$Ti$_2$O$_7$ (\cite{ramirez-nature}). The data show no sign of a phase
transition as would be indicated by a sharp feature in $C(T)$.
Instead, one observes a broad maximum at a temperature
T$_{\rm peak} \sim 1.2$ K, which is of the order of the energy scale of
the magnetic interactions in that material, as measured by the
Curie-Weiss temperature, $\sim1$ K. The specific heat has the appearance of a
Schottky anomaly, the characteristic curve for a system with two energy
levels. At the low temperature side of the
Schottky peak, $C(T)$ falls rapidly towards zero, indicating an almost
complete
freezing of the magnetic moments.

Ramirez {\it et al.} determined the ground state entropy using
a method analogous to that applied by Giauque and co-workers
to water ice.
In general one can only measure
a change in entropy between two temperatures. Giauque {\it et al.}
computed the entropy change of water between liquid helium temperatures
and the gas phase by integrating
the specific heat~(\cite{Giauque}) and then comparing this value
with the absolute entropy calculated for the gas phase
using spectroscopic measurements of
the energy levels of the water molecule.
The difference
gave the residual entropy, later calculated by Pauling~(\cite{Pauling}).
The approach of Ramirez {\it et al.} was to integrate
the magnetic specific heat between
$T_1 = 300$ mK in the frozen regime and $T_2 = 10$ K in the paramagnetic
regime,
where the expected entropy should be $R\ln(2)$ for a two state system.
The magnetic entropy change, $\Delta S$,
was determined by integrating $C(T)/T$
between these two temperatures:
\begin{equation}
\Delta S_{1,2} = \int_{T_1}^{T_2} \frac{C(T)}{T} \;
dT    \;\;\; .
\end{equation}
The lower panel of Fig.3 shows that the
magnetic entropy recovered is approximately 3.9
Jmol$^{-1}$K$^{-1}$, a number that falls considerably short of the
value R$\ln(2)\approx$ 5.76 Jmol$^{-1}$K$^{- 1}$.
The difference, 1.86 Jmol$^{-1}$K$^{- 1}$
is quite close to Pauling's estimate for the
entropy associated with the extensive
degeneracy of ice: $(R/2)\ln(3/2) = 1.68$ Jmol$^{-1}$K$^{-1}$, consistent
with the existence of an ice-rule obeying
spin ice state in Dy$_2$Ti$_2$O$_7$.

 \begin{figure}
\begin{center}
\rotatebox{0}{\includegraphics[width=6cm]{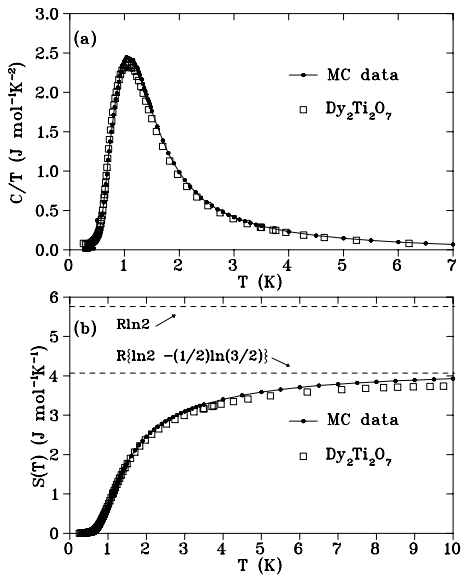}}
\vspace{1mm}
\caption{(A) Specific heat  and (B) entropy data for ${\rm
Dy_{2}Ti_{2}O_{7}}$ 
([29])
compared with Monte Carlo simulation results for the dipolar spin ice
model
([34]),
with $J_{{\rm nn}}=-1.24$ K,  $D_{{\rm nn}} = 2.35$ K and system size $L$ = 4.
}
\end{center}
 \end{figure}

\vspace{-5mm}

\section{Dipolar Spin Ice}

As mentioned above, the magnetic cations Ho$^{3+}$ and Dy$^{3+}$ in
Ho$_2$Ti$_2$O$_7$ and Dy$_2$Ti$_2$O$_7$ carry a very large magnetic
moment, $\mu$, of approximately 10$\mu_{\rm B}$. Furthermore, these
moments are exceedingly well characterized by almost perfect effective
classical Ising spins constrained to point along the local $\langle 111
\rangle$ directions below a temperature of order of 200 K for
Ho$_2$Ti$_2$O$_7$ and 300 K for Dy$_2$Ti$_2$O$_7$.  This is born
out from direct experimental evidence (see Fig.2), magnetization
measurements, inelastic neutron measurements and ab-initio theoretical
calculations~(\cite{Blote,Mamsurova,Bramwell,Jana,Rosenkranz}).
Large magnetic moments are reasonably common among
rare-earth materials, and this gives rise to a sizeable magnetic dipole
energy.  With a cubic unit cell dimension $a\approx 10.1 \AA$, an
estimate of the dipolar energy scale for two $\langle 111 \rangle$
Ising moments, $D_{\rm nn}$, gives:
\begin{equation} D_{\rm nn} = \frac{5}{3}
(\frac{\mu_0}{4\pi})
\frac{\mu^2}{r_{\rm nn}^3} \approx +2.4 \;{\rm K}                  \;\;\;,
\end{equation}
\noindent
where $r_{\rm nn}=(a/4)\sqrt(2)$ is the nearest-neighbor distance.
As discussed below, the 5/3 factor
originates from the orientation of the Ising quantization axes relative
to the vector direction connecting nearest neighbor magnetic moments.
The experimentally determined Curie-Weiss temperature,
$\theta_{\rm CW}$, extrapolated from temperatures below $T \sim 100$ K
 is $\theta_{\rm CW}\sim $+1.9 K for Ho$_2$Ti$_2$O$_7$~(\cite{PRL1})
and +0.5 K for Dy$_2$Ti$_2$O$_7$~(\cite{ramirez-nature}), respectively.
These two values show that $\theta_{\rm CW}$ is of the same order of
magnitude as the nearest neighbor dipolar energy scale $D_{\rm nn}$.
Furthermore, it is well known that rare-earth ions tend to possess very
small exchange energies. Consequently, as originally pointed out by
Siddharthan {\it et al.} (\cite{siddharthan-prl}),
magnetic dipole-dipole interactions dominate exchange in both
Ho$_2$Ti$_2$O$_7$ and Dy$_2$Ti$_2$O$_7$, the opposite
situation to transition
metal compounds where the dipolar forces are typically a very
 weak
perturbation on exchange interactions. We refer to
spin ice materials where magnetic dipolar
interactions are the leading energy interactions
as ``dipolar spin ice'' materials.

In order to consider the combined role of exchange and dipole-dipole
interactions, it is useful to define an effective nearest neighbor
energy scale, $J_{\rm eff}$, for
$\langle 111 \rangle $ Ising spins:
 \begin{equation}
 J_{\rm eff} \; \equiv \;        J_{\rm nn}+D_{\rm nn} \;\;\; ,
\end{equation}
\noindent
where $J_{\rm nn}$ is the nearest neighbor exchange
energy between $\langle 111\rangle $ Ising moments. This simple
description predicts that a $\langle 111\rangle $ Ising system could
display spin ice properties, even for antiferromagnetic nearest
neighbor exchange, $J_{\rm nn}<0$, so long as $J_{\rm eff}=J_{\rm
nn}+D_{\rm nn}>0$.  Fits to experimental data give $J_{\rm nn} \sim
-0.52$ K for Ho$_2$Ti$_2$O$_7$ (\cite{Bramwell-Sq}) and $J_{\rm nn} \sim
-1.24$ K for Dy$_2$Ti$_2$O$_7$ (\cite{hertog}).  Thus, $J_{\rm
eff}$ is positive (using $D_{\rm nn}=2.35$K), hence ferromagnetic and
frustrated, for both Ho$_2$Ti$_2$O$_7$ ($J_{\rm eff} \sim 1.8$ K) and
Dy$_2$Ti$_2$O$_7$ ($J_{\rm eff} \sim 1.1$ K).  It would therefore
appear natural to ascribe the spin ice behavior in both
Ho$_2$Ti$_2$O$_7$ and Dy$_2$Ti$_2$O$_7$ to the positive $J_{\rm eff}$
value as in the simple model of Bramwell and Harris~(\cite{JPCM}).
However, the situation is much more complex than it appears.

Dipole-dipole interactions are ``complicated''; (i) they are
strongly anisotropic since they couple the spin, ${\bf S}_{i}^{\hat
z_{i}}$, and space, ${\bf r}_{ij}$, directions, and (ii) they are also
very long range ($\propto r_{ij}^{-3}$). For example, the second
nearest neighbor distance is $\sqrt 3$ times larger than the nearest
neighbor distance, and one then has a second nearest neighbor energy scale,
$D_{\rm nnn} \sim 0.2D_{\rm nn}$. This implies an important perturbation
compared to $J_{\rm eff}=J_{\rm nn}+D_{\rm nn} < D_{\rm nn}$, especially in the
context of antiferromagnetic (negative) $J_{\rm nn}$.  Specifically, for
Dy$_2$Ti$_2$O$_7$, the second nearest neighbor energy scale is about 40\%
of the effective nearest neighbor energy scale, $J_{\rm eff}$ $-$
a large proportion!  One might therefore expect that the
dipolar interactions beyond nearest neighbor would
cause the different ice-rule states to have
different energies, hence possibly breaking the degeneracy
of the spin ice manifold. This would result in
long range
N\'eel order with a critical temperature $T_N \sim O(D_{\rm nn})$.
Thus, there arises a puzzling and fascinating problem posed by the
dipolar spin ice
materials, that we summarise in two questions:

\begin{itemize}
\item { Are the experimental observations of spin ice behavior
consistent with dominant long range dipolar interactions?}
\item {
If so, then why do long range dipolar interactions fail to
destroy spin ice behavior~?}
\end{itemize}

The minimal model one needs to consider to investigate these
questions
includes nearest neighbor exchange and long range magnetic dipole
interactions:
\begin{eqnarray} \label{hamiltonian} H&=&-J\sum_{\langle
ij\rangle}{\bf S}_{i}^{{\hat z}_{i}}\cdot{\bf S}_{j}^{ {\hat z}_{j}}
\nonumber \\ &+& Dr_{{\rm nn}}^{3}\sum_{i>j}\frac{{\bf S}_{i}^{{\hat
z}_{i}}\cdot{\bf S}_{j}^{{\hat z}_{j}}}{|{\bf r}_{ij}|^{3}} -
\frac{3({\bf S}_{i}^{{\hat
 z}_{i}}\cdot{\bf r}_{i j}) ({\bf S}_{j}^{{\hat z}_{j}}\cdot{\bf
r}_{ij})}{|{\bf r}_{ij}|^{5}} \;\; .  \end{eqnarray}
The first term is
the near neighbor exchange interaction, and the second term is the
dipolar coupling between the $\langle 111 \rangle)$ Ising magnetic moments.
For the open pyrochlore lattice structure, we expect
further neighbor exchange coupling to be very
small~(\cite{greedan-open-pyro}), so it can be neglected
to a good approximation.
Here the spin vector ${\bf S}_{i}^{{\hat
z}_{i}}$ labels the Ising moment of magnitude
$\vert {\bf S}_{i}^{{\hat z}_{i}} \vert=1$ at lattice site $i$ and
oriented along the local Ising $\langle 111 \rangle$
axis ${{\hat z}_{i}}$. The
distance $\vert {\bf r}_{ij}\vert$ is measured in units of the nearest
neighbor distance, $r_{\rm nn}$. $J$ represents the exchange
energy and $D=(\mu_{0}/4\pi)\mu^{2}/r_{\rm nn}^{3}$. Because of the
local Ising axes, the effective nearest neighbor energy scales are
$J_{\rm nn}\equiv J/3$
and, as mentioned above, $D_{\rm nn}\equiv
5D/3$, since ${\hat z}_i\cdot{\hat z}_j = -1/3$ and $({\hat
z}_i\cdot{\bf r}_{ij})  ({\bf r}_{ij}\cdot{\hat z}_j) = -2/3$.
If $D_{\rm nn} = 0$ one obtains the spin ice model originally proposed
by Harris et al (\cite{PRL1,JPCM}), henceforth referred to as
the ``near neighbour spin ice model''.

Siddharthan {\it et al.} (\cite{siddharthan-prl})
first addressed the role of dipole-dipole
interactions in Ising pyrochlore materials, both Dy$_2$Ti$_2$O$_7$
(\cite{ramirez-nature}) and
Ho$_2$Ti$_2$O$_7$~(\cite{siddharthan-prl,siddharthan-prb}).
They considered Eq.~\ref{hamiltonian} above, but restricted
the dipolar lattice sum up to the first five nearest
neighbors (\cite{siddharthan-prl}), or ten and twelve nearest
neighbors~(\cite{siddharthan-prb}).  With the values they used for
$J_{\rm nn}$ and $D_{\rm nn}$ the conclusion that they reached was
that a spin ice state could exist for their model of
Dy$_2$Ti$_2$O$_7$ but not for their model of Ho$_2$Ti$_2$O$_7$,
where a partially ordered state was predicted.
It was suggested
that a transition to this partially ordered state was
consistent with a sharp rise in the experimental specific heat
of Ho$_2$Ti$_2$O$_7$ observed below $\sim 1$ K.
The reported specific heat rise has more
recently been shown to be consistent with
the freezing of the nuclear spins of $^{165}$Ho, as discussed below.
However, the results of Ref.~(\cite{siddharthan-prl})
did appear to give a negative answer to the first of the
above questions: the dipolar model, it seemed, was
inconsistent with the experimental data that
supported a disordered spin ice ground state
for Ho$_2$Ti$_2$O$_7$~(\cite{PRL1}).

On the other hand, direct dipolar lattice sums are
notoriously tricky to
handle, and truncation of dipolar interactions at some arbitrarily
chosen fixed finite distance often gives rise to spurious
results. One approach that has been
successfully used to obtain reliable quantitative results for real
dipolar materials (\cite{White}) is the Ewald
summation method (\cite{Ewald}). This method is conceptually
akin to the well known Madelung approach used to calculate
electrostatic Coulomb energies in ionic crystals.  The
Ewald summation method generates an absolutely convergent effective
dipole-dipole interaction between two spins $i$ and $j$. This is done
by periodically replicating a specified volume, and summing
convergently the interactions between $i$ and $j$ and all the
periodically repeated images of $j$.
Once an
effective dipole-dipole interaction between spins $i$ and $j$ within
the simulation cell has been derived via the Ewald summation
technique, one can perform Monte Carlo simulations using the standard
Metropolis algorithm. This technique was applied to the
dipolar spin ice model by
den Hertog and Gingras (\cite{hertog}), who found,
contrary to the results of refs.(\cite{siddharthan-prl,siddharthan-prb}),
no sign of full or partial ordering for all $J_{\rm eff}/D_{\rm nn}
\ge 0.09$. The system was characterized as having spin-ice behavior by
determining the entropy, via numerical integration of $C(T)/T$.  For
$J_{\rm eff}/D_{nn} \ge 0.09$ the recovered magnetic entropy
was found to be within a few per cent of Pauling's
value $R[\ln(2)-(1/2)\ln(3/2)]$ (\cite{hertog}).  It was shown
in Ref.~(\cite{hertog}) that the only free parameter
in the theory is $J_{\rm
nn}$, which can be determined by comparing the experimental and
theoretical specific heat temperature peak $T_{\rm peak}$ referred to
above.
This procedure gives $J_{\rm nn}=-1.24$ K for Dy$_2$Ti$_2$O$_7$
(\cite{hertog}) and $J_{\rm nn}=-0.52$ K for Ho$_2$Ti$_2$O$_7$
(\cite{Bramwell-Sq}).  Figure 3 shows the very good agreement
obtained from Monte Carlo simulations results, using Ewald summation
methods~(\cite{hertog}), with the experimental results for
Dy$_2$Ti$_2$O$_7$ (\cite{ramirez-nature}).
These numerical results give a definite
answer to at least the first of the two questions
posed above:
dominant long ranged dipolar interactions are
indeed consistent with the spin ice behaviour observed
in the dipolar spin ice materials~(\cite{note}).
We find it remarkable that long range dipolar coupling can,
through some effective self-screening, restore the ice-rules
degeneracy to such a striking degree (\cite{Gingras}). Consequently, we
feel that the second
question $-$ the ``why?'' $-$ has not yet
been answered in any simple and definite way.

In order to further test the proposal that the formation of the
spin ice state in real materials is due to long range
dipole-dipole
interactions, one needs to compare the spin-spin correlations in real
systems with that predicted by the dipolar spin ice model. Recently,
elastic neutron scattering experiments on a flux grown single crystal
of Ho$_2$Ti$_2$O$_7$ have found excellent agreement with the predictions of
the dipolar spin ice model, and establish unambiguously the spin ice
nature of the zero field spin correlations in that
material~(\cite{Bramwell-Sq}). These results also show that the
dipolar interactions beyond nearest neighbour do slightly favour some
of the spin ice states over others although they do not
significantly affect the ground state entropy.

Figure 4 shows the elastic neutron scattering pattern of
Ho$_2$Ti$_2$O$_7$ at $T\sim 50$ mK and compares it with theoretical
predictions for the near neighbour and dipolar spin ice models.
The pattern for near neighbour spin ice successfully
reproduces the main features of the experimental pattern,
but there are important differences, both qualitative and quantitative,
notably in the
extension of the $0,0,0$ intense region
along $[hhh]$ and the relative intensities of the regions around
$0,0,3$ and $3/2,3/2,3/2$. Also, the experimental data shows much
broader regions of scattering along the diagonal directions.
The dipolar model successfully accounts for these discrepencies.
In particular, it predicts the four intense regions around
$0,0,0$, the relative intensities of the regions around $0,0,3$ and
$3/2,3/2,3/2$ and the spread of the broad features along the diagonal.
The neutron scattering data can in fact be accurately and quantitatively
accounted for by the dipolar model with no free parameter
once $J_{nn}$ has been determined by the height
of the specific heat peak~(\cite{Bramwell-Sq}).
Qualitatively similar scattering has been observed in water ice
and described by an ice-rules configuration of protons~(\cite{Li}).

\begin{figure}
\label{fig1}
\rotatebox{0}{\includegraphics[height=18.5cm]{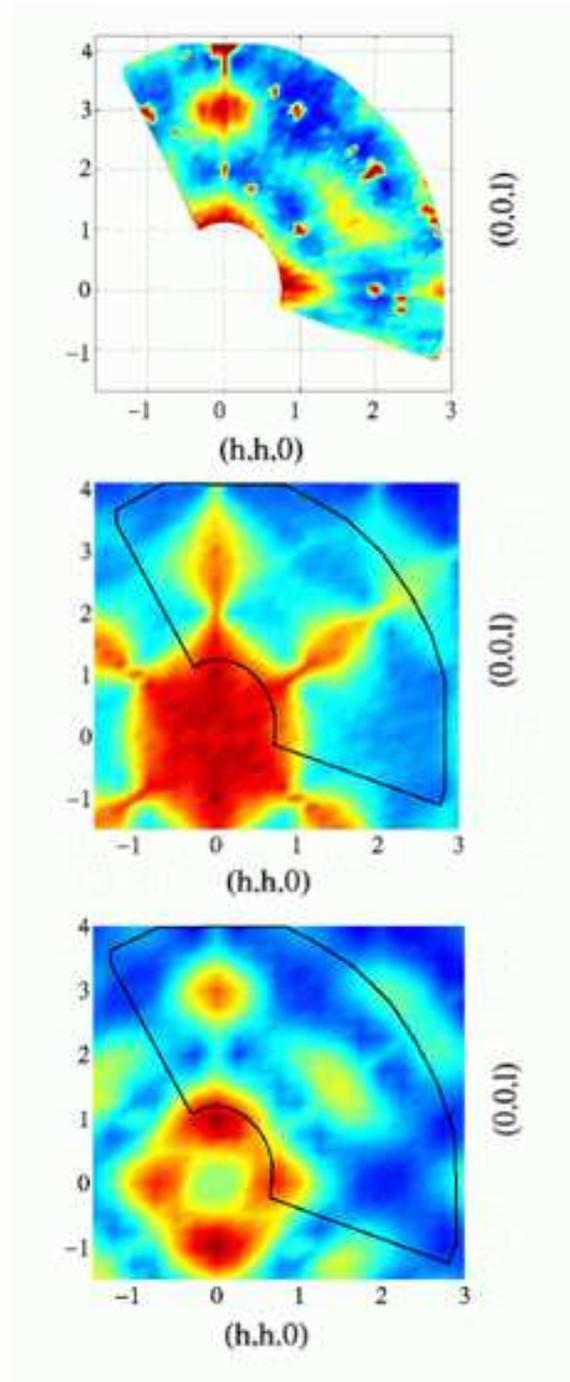}}
\caption{(A)
Experimental neutron scattering pattern of ${\rm Ho_{2}Ti_{2}O_{7}}$ in
the $(hhl)$ plane of reciprocal space at $T\sim 50$
mK
 ([28]).
 Dark blue
shows the lowest intensity level, red-brown the highest. Temperature
dependent measurements have shown that the sharp diffraction spots in
the experimental pattern are nuclear Bragg peaks with no magnetic
component.  (B) Calculated neutron
scattering for the nearest neighbor spin ice model at
$T=0.15J$.  (C) Calculated neutron scattering
for the dipolar spin ice model at $T = 0.6$ K. This can be compared with
the experimental scattering because the latter is
temperature-independent in this range. The areas defined by
the solid lines denote the experimental data region of (A).
}
\end{figure}

To complete the description of the static
properties of Ho$_2$Ti$_2$O$_7$ it was shown in Ref.~(\cite{Bramwell-Sq})
that the specific heat could be very accurately described by the
sum of the dipolar spin ice contribution, with
$D_{\rm nn}  = 2.35$ K, $J_{\rm nn} = -0.52$ K, and a nuclear
spin contribution with level splitting $\sim 0.3$ K,
the large value typical of a Ho$^{3+}$ salt. Analysis of
the hyperfine contribution followed the early work of
Bl\"ote {\it et al.} (\cite{Blote}), who observed that the specific heat of
isostructural Ho$_2$GaSbO$_7$ can be accurately fitted by the
sum of two Schottky contributions, one arising
from the nuclear and one from the electronic spins.
It is interesting to note that these
authors had also commented on some evidence for a residual entropy in
Dy$_2$Ti$_2$O$_7$, later attributed by Ramirez {\it et al.} to spin
ice behaviour~(\cite{ramirez-nature}).

The above results show that dipole-dipole interactions can cause spin ice
behavior, and that the simple spin Hamiltonian defined in Eq.~\ref{hamiltonian}
can provide a
quantitative description of experimental results on real materials. In the
next section we discuss how, if dynamics can be preserved in
simulations, that dipole-dipole interactions do
stabilize a long range N\'eel order at a critical temperature $T_c
\ll D_{\rm nn}$, hence partially addressing the second question above.

\section{Open Issues and Avenues for Future Research}

Among open issues in the physics of dipolar spin ice materials are
the question of the ``true ground state'', the
magnetic field
dependent behaviour,
the effect of diamagnetic dilution, the nature of the spin dynamics, and the
properties of spin ice related materials.

The question of a ``true ground state'' has long intrigued researchers on
water ice, and the same question applies to spin ice.  A common
interpretation of the ``third law'' of thermodynamics is that the true
ground state of a real system must be ordered, without entropy. If the
system is ergodic, that is, it can explore all its possible arrangements,
then presumably it should settle into its absolutely minimum energy ground
state, which we refer to as its ``true'' ground state. This is not
observed, either in water ice or in the spin ice materials.  However, the
experimental zero point entropy does not necessarily
mean that the system does
explore its spin ice manifold on the time scale of the experiment.
Rather, it suggests that the system, with a finite correlation length, is
self-averaging, so the thermodynamic average over one state is
equivalent to the canonical average over an ensemble of states
(\cite{Peter}). If the system is ``stuck'' in a disordered state then, as
discussed in the last section, the
following question arises:  would the long range
part of dipolar interaction stabilise
a ``true ground state'' of lower energy than all the others spin ice states
if it was not dynamically inhibited from forming as the system is cooled
through the temperature $ T \sim J_{\rm eff}$
at which the ice-rule manifold develops?
Recent theoretical work on the
dipolar spin ice model has answered the above
question in the affirmative:
the low-energy frozen state that forms does indeed depend
on the dynamics introduced~(\cite{Melko}).
Numerical simulation of the dipolar spin ice model using single spin
flip gives a disordered ground state as observed in experiment
(\cite{PRL1,Fennell,Kadowaki}), but the introduction of ``loop moves'',
that is, correlated flipping of extended groups of spins (Fig. 1C), gives
rise to a first order transition to an ordered state.
 The transition
is predicted to be independent of $J_{\rm nn}$, with $T_c\approx
0.077D_{\rm nn}
\sim 0.18$ K for both Ho$_2$Ti$_2$O$_7$ and Dy$_2$Ti$_2$O$_7$~(\cite{Melko}).
The ordered state is
illustrated in Fig. 5. It corresponds to a tetrahedral ``two in $-$ two
out'' basis of spins, inverted by two out of the three face centering
translations
in the unit cell. Interestingly, a structure of related symmetry (the
``$Q = X$'' phase) has been observed in Ho$_2$Ti$_2$O$_7$ after the application
of a field, as discussed below.  An intriguing possibility is that the
predicted ordered state might be accessible by a suitable field and
temperature cycle in the spin ice regime.  The theoretical ordered state
identified
in (\cite{Melko}) is compatible with the critical (soft) mode identified
in a mean-field theory treatment of the dipolar spin ice model
(\cite{Kadowaki,Gingras}).
It would appear possible that it is the true ground state of the real
materials, as the Ewald method of (\cite{hertog,Melko}) gives the correct
quantitative
description of
experiment in regard to the paramagnetic disordered spin ice state. However,
refs.~(\cite{siddharthan-prl,siddharthan-prb}) argue that it is the
truncation method, rather than the Ewald method, that gives the true ground
state of the dipolar model, suggesting that the mathematical problem of
determining the true ground state is not yet resolved~(\cite{note}).

\vspace{-3cm}
 \begin{figure}
\begin{center}
\rotatebox{-90}{\includegraphics[width=8cm]{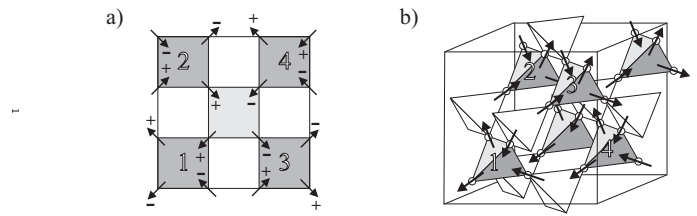}}
\vspace{-2cm}
\caption{The predicted long range ordered state of dipolar spin ice.
Projected down the $z$
axis (A), the four tetrahedra making up the cubic unit cell appear as
dark grey squares.  The light grey square in the middle does not
represent a tetrahedron; however its diagonally opposing spins occur in
the same plane.  The component of each spin parallel to the $z$ axis is
indicated by a $+$ and $-$ sign.  In perspective (B), the four
tetrahedra of the unit cell are numbered to enable comparison with
(A).
}
\end{center}
 \end{figure}

The zero field static properties of the spin ice materials
Ho$_2$Ti$_2$O$_7$ (\cite{Bramwell-Sq})  and
Dy$_2$Ti$_2$O$_7$~(\cite{ramirez-nature,hertog})
and most recently Ho$_2$Sn$_2$O$_7$ (\cite{Kadowaki}) can now be
considered to be very well described by the dipolar spin ice model. The
essential scenario is a continuous freezing process into a disordered
magnetic state below $\sim 1$ K. In the earliest study of
Ho$_2$Ti$_2$O$_7$ it was pointed out that its field dependent static
properties were far from trivial~(\cite{PRL1}). Two ordered magnetic phases
were
observed to rapidly develop in a magnetic field of $\sim 0.1$ T applied
along $[1\bar{1}0]$ at base temperature $\sim 50$ mK. Both of these can be
described with the same tetrahedral basis as the face centered cubic
crystal structure, but one has ``ferromagnetic'' face centering, the $Q = 0$
phase, and the other has partial ``antiferromagnetic'' face centering, the
$Q = X$ phase.  It has been shown that both states fulfill
the spin ice rules (\cite{PRL1}). The field induced order displays
strong history dependence as a function of field (up to 3 T) and
temperature (50 mK - 2 K). Low temperature susceptibility studies,
discussed below, have confirmed that a strongly history dependent
magnetic moment can be induced by an applied field below
a temperature of $\sim 0.7$ K
(\cite{Matsuhira}) (Fig. 6).
Theoretical studies of the near neighbour
spin ice model~(\cite{PRL2}) suggests several interesting effects as the
field is applied along the other main symmetry directions, $[100]$ and
$[111]$. Application of a field along $[100]$ breaks symmetry such that
each tetrahedron has the same ``two in $-$ two out'' state, but a symmetry
sustaining first order phase transition is predicted, analogous to the
liquid gas transition or those that can be observed in ferroelectrics
(\cite{PRL2,ferro}).  A field of several Tesla along $[111]$ breaks
the ice rules to give a state with ``three in $-$ one out''  (or
vice-versa). This transition, also predicted by the dipolar
model~(\cite{Melko_thesis}), has been observed by single crystal
magnetization measurements at $2 K$, first on ${\rm Ho_2Ti_2O_7}$ by
Cornelius and Gardner~(\cite{Cornelius}) and more recently on ${\rm
Dy_2Ti_2O_7}$ by Fukuzawa {\it et al.}~(\cite{Fukazawa}). These two
studies reveal a potentially
interesting difference between the two compounds: while the saturation
magnetization of ${\rm Dy_2Ti_2O_7}$ along the three main symmetry
directions is in remarkable quantitative agreement with the predictions of
the spin ice rules~(\cite{Fukazawa}), that for ${\rm Ho_2Ti_2O_7}$ shows a
significant departure~(\cite{Cornelius}). At $2 K$ the field-dependent
behaviour of the spin ice materials appears to be equally well-described by
the dipolar and near neighbour models. However, it is most unlikely that
the success of the oversimplified near neighbor model will extend to lower
temperatures. Hints of this have already been
observed in the specific heat of Dy$_2$Ti$_2$O$_7$, where
polycrystalline samples display several thermal anomalies as a function
of field and temperature which do not agree with the near neighbour
model~(\cite{ramirez-nature}). It would be hoped that the dipolar spin ice
model
will account for these details~(\cite{Melko_thesis}).
The field dependent and related
field induced relaxational dynamic behavior of spin
ice materials promises to be an interesting area of research. For example,
recent neutron studies show that the magnetization
process in the spin ice regime of a single crystal sample of
Dy$_2$Ti$_2$O$_7$ occurs via a series of steps and
plateaus~(\cite{Fennell}).

 \begin{figure}
\begin{center}
\rotatebox{-90}{\includegraphics[width=8cm]{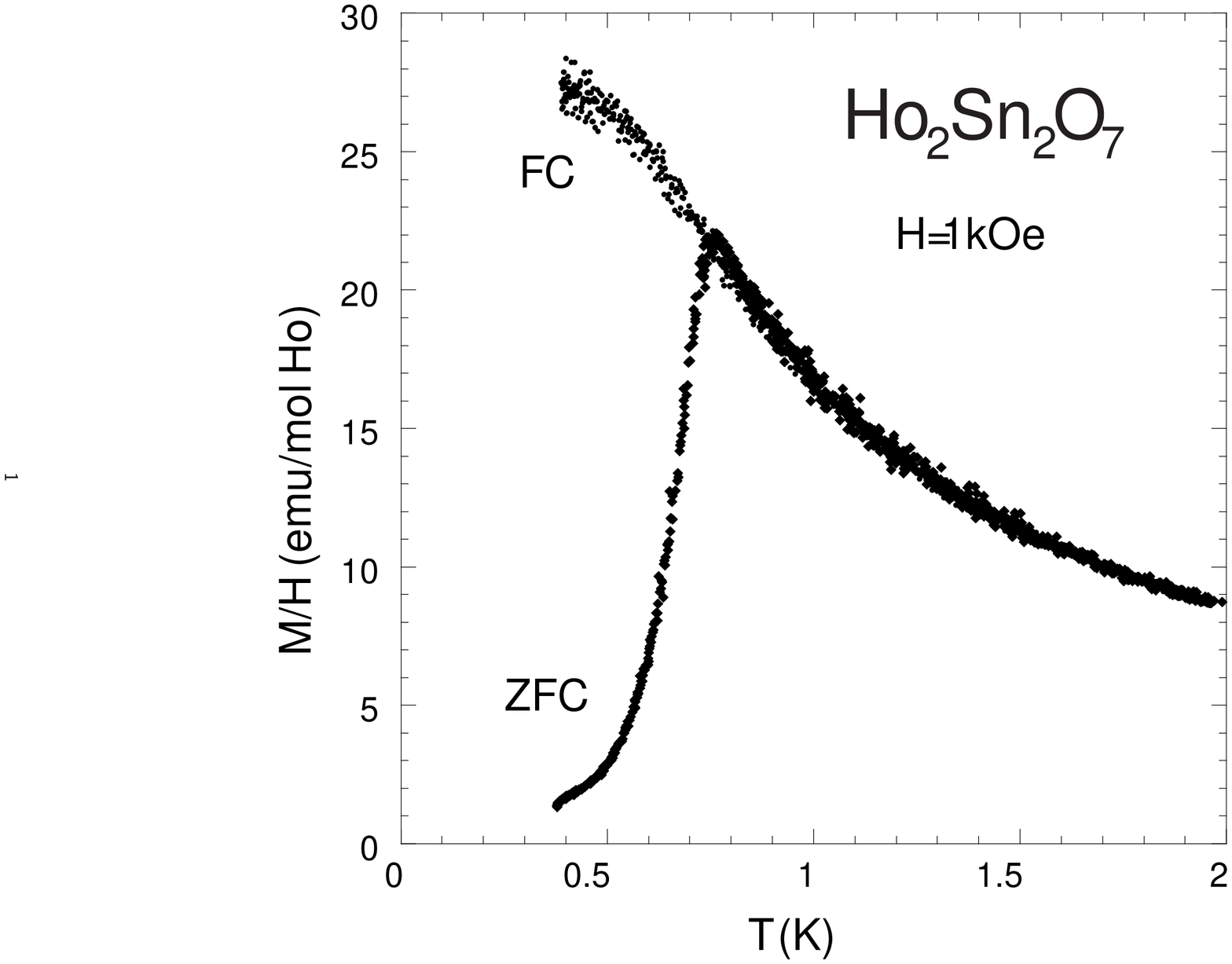}}
\caption{
Temperature dependence of the dc magnetization of the Ho$_2$Sn$_2$O$_7$
spin ice materials in an applied magnetic field, $H$, of 1kOe
 ([49]).
Zero field cooled, ZFC, denotes a
procedure that involves cooling in zero field and measuring the magnetization,
$M$,
upon warming with the field $H$ on.  Field cooled, FC,
involved measuring the magnetization, $M$, upon cooling with
the field $H$ on.
}
\end{center}
 \end{figure}

The dynamics of the Ising spins in spin ice materials is of
broad potential interest in context of the long standing problem
of the proton dynamics in the various phases of water
ice~(\cite{icebook}). Matsuhira {\it et al.}
have investigated the dynamic properties
of the spin ice materials Ho$_2$Ti$_2$O$_7$ and Ho$_2$Sn$_2$O$_7$
by measuring both dc and ac magnetization~(\cite{Matsuhira}). History
dependence of the dc magnetization in the spin ice phase shows
a sharp splitting between field cooled and
zero field cooled susceptibility, indicative of the expected spin freezing
process (Fig. 6). The characteristic
response of the ac magnetization is peaked near $\sim 1$ K which shifts to
higher temperature
with increasing frequency.  This can be analysed in terms of an
thermally activated Arrhenius-type spin relaxation with a characteristic
relaxation time of $\sim10^{-10}$ s and an activation barrier of 28 K and
20 K for
Ho$_2$Ti$_2$O$_7$ and Ho$_2$Sn$_2$O$_7$, respectively. This behaviour may
be compared with that of other types of magnetic system that exhibit spin
freezing~(\cite{Mydosh}).
In fine particle magnets one observes the thermally activated flipping of
independent spin clusters which becomes frozen, or ``blocked'', at low
temperature, while in certain dilute transition metal alloys (such as
AuFe$_{0.01}$) one observes ``spin glass'' behaviour in which the spin
freezing is a cooperative process.
The behaviour of Ho$_2$Ti$_2$O$_7$ and Ho$_2$Sn$_2$O$_7$ appears
more similar to a blocking phenomenon than to a spin glass transition
~(\cite{Matsuhira}).
Very recently, two ac-magnetization studies of Dy$_2$Ti$_2$O$_7$ have been
reported~(\cite{Matsu2,Snyder}). The first of these studies, by Matsuhira
{\it et al.}~(\cite{Matsu2}) finds behaviour below 2 K analogous to
Ho$_2$Ti$_2$O$_7$ and Ho$_2$Sn$_2$O$_7$, which, if fitted to an Arrhenius
expression gives an activation barrier of 10 K. This process is cautiously
ascribed to spin ice freezing. In addition, Matsuhira {\it et al.} report
an ac-susceptibilty peak that is observed above 10 K, which indicates an
Arrhenius-type response with a large activation energy of 220 K. This high
temperature susceptibility peak is also observed in the second study, by
Snyder {\it et al.}~(\cite{Snyder}). However, the two reports disagree in
their interpretation of this feature.
Snyder {\it et al.} argue that the high temperature susceptibility peak
reflects a single relaxation time and ascribe this to spin ice freezing.
Matsuhira {\it et al.}, in contrast, suggest that the high temperature
susceptibility peak indicates a spread of relaxation times, at least near
to $T \sim 10$ K, but do not speculate on the physical origin. It would be
premature to discuss which interpretation might be correct, but it would
seem to us that in view of the energy scale, another possible physical
origin worth investigating is
spin-lattice relaxation. The dynamics of spin ice materials is clearly an
active area of research. We also note $\mu$SR studies on Ho$_2$Ti$_2$O$_7$
~(\cite{HarrisMMM}) and theoretical investigations into ordered state
selection in spin ice materials via quantum fluctuations, as ongoing
reseacrh activities~(\cite{Moessner}).

Possible modifications of spin ice materials include doping with
diamagnetic impurities. In water ice, doping with KOH (which
effectively removes protons) leads to the ordered phase known as ice
XI~(\cite{Tajima}): it is believed that the holes introduced in the proton
structure
``free up'' the dynamics so that the system can find its true ordered
ground state. In the case of spin ice it is not clear that dilution with
diamagnetic impurities, such as replacing Ho$^{3+}$ and Dy$^{3+}$ by
non-magnetic Y$^{3+}$,  will have the same effect~(\cite{Melko_thesis}).
However,
dilution is likely to have
some effect on the dynamical properties on the spin ice state.
With considerable dilution, one would expect to form a spin glass state.
Such studies might go a long way to clarifying the precise
differences between the spin ice state and the spin glass state, and their
similarities, or lack thereof, in their freezing mechanisms. The fact that
spin ice is ``self-averaging'' (see above) while a spin glass is
not~(\cite{Mydosh}) points to the essentially simpler nature of the spin
ice freezing process. However it is not yet clear how such a detailed
difference will manifest
itself experimentally.

Perhaps the most fruitful area of future research will be into
materials and models which can be considered derived from or related to
spin ice. At a theoretical level the near neighbour spin ice model
of~(\cite{JPCM}) is a ``sixteen vertex'' model of the ferroelectric type long
studied by theoreticians~(\cite{Lieb}); the discovery of spin ice materials
is giving a new experimental relevance to these models (\cite{Watson,Huber}).
Modifications of the
near neighbour model may be designed to include quantum mechanical
effects~(\cite{Moessner}) or the effect of finite
anisotropy~(\cite{Champion}). Such models might be relevant to as-yet
undiscovered spin ice materials based on transition metal ions
where spin values are smaller and exchange is larger.
Known spin ice related materials
include the frustrated pyrochlore magnet
Tb$_2$Ti$_2$O$_7$, the behaviour of which remains essentially
mysterious~(\cite{tb2ti2o7-prl,tb2ti2o7-prb}).
Tb$_2$Ti$_2$O$_7$
has been described as a ``cooperative paramagnet'' (\cite{tb2ti2o7-prl}),
but it may also share some properties with the spin ice
materials~(\cite{hertog,tb2ti2o7-prb}).
Another
example is Nd$_2$Mo$_2$O$_7$, in which spin-ice like correlations on
the Nd sublattice perturb the metallic behaviour on the Mo sublattice
to give a striking anomalous Hall effect~(\cite{nd2mo2o7}). A final example is
Dy$_2$Ir$_2$O$_7$, a metallic material, but in which the Dy spins appear
to be antiferromagnetically coupled, leading to an ordering
transition~(\cite{Maeno}).  The possiblity that one could obtain a
metallic spin
ice
is an intruiging one:  the conduction electron mediated coupling of
localised spins, the so-called Ruderman-Kittel-Kasuya-Yosida (RKKY)
interaction, is similarly
long-ranged to the dipolar interaction.  Perhaps this long range
coupling could, like the dipolar interaction, stabilise spin ice
behaviour. If so, it would represent another novel realisation of
concept of spin ice.

In conclusion,
the previous discussion emphasises the current interest in systems in which
the
strong magnetic anisotropy of localised $f$-electrons
leads to novel
collective effects based on frustration. The spin ice phenomenon
exemplifies the richness of the intrinsic physics of frustration while, at the
same time,
affording a close connection between theory and experiment.
This has allowed real progress to be be made and some
surprising properties (for example, the effective self-screening of the long
range
dipolar interactions) to be identified. One may hope that the current
experimental and theoretical studies of spin ice materials will lead to a
deeper
understanding
of the effects of frustration in related rare-earth
(\cite{nd2mo2o7,gd2ti2o7,Palmer})
and transition metal magnets. This in turn should provide
a solid platform from which we can consolidate our
understanding of the diverse electronic phenomena
in which frustration is, or might be, implicated.
Exciting topical examples include heavy-fermion behavior in LiV$_2$O$_4$
(\cite{liv2o4-Kondo,liv2o4-Lee}),
spin-Peierls related phenomenon in ZnCr$_2$O$_4$ (\cite{zncr2o4}), and
superconductivity in
Cd$_2$Re$_2$O$_7$ (\cite{cd2re2o7-Jin,cd2re2o7-Hanawa,cd2re2o7-Sakai}).
We expect that
new collective phenomena in spin ice materials and their relatives will
continue to capture the attention of condensed matter physicists and
solid state chemists for years to come~(\cite{ack}).

\end{document}